# Fundamentals of Gas Micropattern Detectors

V. Peskov[1,*], P. Fonte[2,3], M. Danielsson[1], C. Iacobaeus[4],
J. Ostling[1,4], M. Wallmark[1,4]

1-Royal Institute of Technology, Stockholm, Sweden
2-LIP, Coimbra, Portugal
3-ISEC, Coimbra, Portugal
4-Karolinska Institute, Stockholm, Sweden

## Abstract

We performed a new series of systematic studies of gain and rate characteristics of several micropattern gaseous detectors. Extending earlier studies, these measurements were done at various pressures, gas mixtures, at a wide range of primary charges and also when the whole area of the detectors was irradiated with a high intensity x-ray beam.

Several new effects were discovered, common to all tested detectors, which define fundamental limits of operation. The results of these studies allow us to identify several concrete ways of improving the performance of micropattern detectors and to suggest that in some applications RPCs may constitute a valid alternative. Being protected from damaging discharges by the resistive electrodes, these detectors feature high gain, high rate capability ($10^5$ Hz/mm$^2$), good position resolution (better than 30 μm) and excellent timing (50 ps σ).

[*] Corresponding author: V. Peskov, Physics Department, Royal Institute of Technology, Frescativagen 24, Stockholm 10405, Sweden (telephone: 468161000, e-mail: vladimir.peskov@cern.ch).

## I. INTRODUCTION

In the last decade there was a chain of inventions of new micropattern gaseous detectors: MSGC, CAT, GEM, MICROMEGAS and many others (see [1], [2]). Due to their promising properties, especially a potential capability for an excellent position resolution, they were almost immediately adopted as tracker devices for some large-scale experiments at CERN and elsewhere. However, as it was later discovered, all micropattern gaseous detectors suffer from two main problems: the maximum achievable gain drops with the counting rate and in the presence of heavily ionizing particles [3]-[5]. The recent experience of the CMS and HERA-B tracker detectors shows that one should take these problems very seriously [6]-[7].

Extending our earlier studies, in this study we performed further systematic measurements of the gain characteristics of micropattern gaseous detectors as a function of the amount of primary charge, counting rate, number of amplification steps and gas pressure, in order to obtain some strategic guide for their improvement.

## II. EXPERIMENTAL SETUP

The experimental setup (Fig. 1) was similar to the one described in [3], [4], [8], [9], except for two modifications: elevated pressures, up to 10 atm, were possible and a much more powerful x-ray gun was available, allowing to expose the detectors to counting rates larger than $10^5$ Hz/mm$^2$ over the whole active area of 10 cm $\times$ 10 cm.

The following micropattern detectors were tested: MSGC, MICROMEGAS, GEM, CAT, MICRODOT and glass capillary plate [10], as well as a set of parallel-plate avalanche chambers (PPAC) and resistive plate chambers (RPC) with gaps varying from 0.1 to 3 mm. The tests were done with single or few-step configurations.

The MSGC's used were manufactured on Desag glass with electrodes made of chromium and placed at 0.2, 1, 3 or 5 mm pitches. All anode strips were 10 μm wide.

Several MICROMEGAS designs were tested: a commercial design had an anode plate made of printed circuit board with copper strips at a pitch of 300 μm; the others were manufactured by us. Their anode plates were made from ceramic or Si plates covered lithographically by Al or Cr strips 50 μm pitch. The cathode was a stretched Ni mesh 12μm pitch supported by fishing lines at 2mm pitch, following [11].

The GEMs were obtained from CERN and had a rather standard geometry: holes of 100 μm diameter at 140 μm pitch.

The CAT was constructed from a GEM whose the anode was kept in contact with a metallic plate.

MICRODOT detectors were custom-made on G10 board with anode and cathode diameters of 30 μm and 300 μm, respectively.

The capillary plates had a thickness of 0.25 to 0.4 mm and a capillary diameter of 50 or 100 μm. The plates were treated with hydrogen to decrease their resistivity [10].

PPACs were tested with three main designs of the cathode electrode: 1) metallic mesh; 2) flat, well polished, metallic sheet; 3) glass and ceramic plates with an evaporated metallic layer.

In the case of RPCs also two main cathode designs were tested: 1) metallic mesh or 2) well polished sheet made of semiconductive materials. For these we used Pestov glass ($10^{10}$-$10^{11}$ $\Omega\cdot$cm), ceramics and n or p-type silicon with different doping levels (resistivity 0.01-2000Ω·cm). The anode plates of the RPCs and of some custom-made MICROMEGAS were also manufactured from Pestov glass or Si and were lithographically covered with aluminium or chromium strips at 30 to 50 µm pitch. Similar plates were also used for the position-sensitive readout of GEMs and capillary tubes, being the strips connected to charge-sensitive amplifiers.

The gap between the upper drift electrode and the detector varied from 3 to 11 mm and the transfer gap between detectors in multistep configuration varied from 1 to 3 mm.

For position resolution measurements a vertical slit 20µm width was used attached to a table, which could be moved with micrometric accuracy

## III. Results at 1 Atmosphere

### A. Very Low Rate

It was established earlier [3] that at very low counting rates, below to 1 Hz/mm$^2$, the maximum achievable charge (MAC - the amount of total charge in the avalanche at which breakdown appears) of micropattern detectors is determined by the following limit

$$An_0 < 10^8 \textit{ electrons} \qquad (1)$$

where $A$ is the gas gain and $n_0$ is the number of primary electrons. Measurements presented in the present study confirm that in general this statement is correct, however revealing some important details. Note that this limit coincides with so called Raether limit established earlier for large-gap PPAC.

It should be understood, however, that in practical situations, to avoid an excessive number of potentially damaging sparks, one has to work at gas gains one or two orders of magnitude smaller than the gains that correspond to the MACs presented here.

As an example, Fig. 2 shows the typical dependence of the MAC on $n_0$. One can see that the Raether limit is reached only for PPACs and $n_0 > N_{crit}$, with $N_{crit} \approx 10^3$ electrons. In this situation, for quencher concentrations larger than a few percent, the MAC only slightly depends on gas mixture.

For $n_0 < N_{crit}$ the MAC can be orders of magnitude smaller than the Raether limit. At these values of $n_0$ the MAC may strongly depend on gas mixture. Moreover, it is typical of micropattern detectors that at small $n_0$ the MAC achieves a maximum in gas mixtures having a large value of $d(\ln A)/dV$, where $V$ is the applied voltage (see [8], [9], [12] for more details). These are usually helium and neon based mixtures with a small concentration of a quencher gas.

Qualitatively the same dependence of the MAC vs. $n_0$ was found for all tested detectors.

Fig. 3 shows the typical dependence of the MAC on the gap thickness for several detector types, with $n_0 \geq N_{crit}$. The MAC increases almost linearly with the gap width, leading to larger values for the thick gap gaseous detectors (wire or PPAC type) when compared to the micropattern detectors. Actually, the classical Raether limit (1) is only met by thick-gap PPACs. Earlier results concerning this type of detector can be also found in [13].

### *B. Rate Effect*

It was discovered earlier that the MAC of micropattern detectors drops with counting rate (see for example [3] and references therein).

Two hypotheses have been suggested so far to explain this effect. One is based on the possible contribution of regions with higher local electric fields near the cathode (for example, cathode strips of MSGC or cathode edges near GEM's walls) to the overall multiplication process [14]. The other one is based on the "cathode excitation effect": emission of jets of electrons from the cathode, bombarded by positive ions [4], [15].

To address this problem we performed MAC vs. counting rate measurements in several PPACs designs that differ only in the nature of the cathode electrode. In one case the cathode was manufactured from wire mesh and in the other cases from a well-polished metallic plate and glass and ceramic plates with vacuum evaporated metallic layers. In the first case the electric field lines were concentrated near wires, while in the second case they were uniform. Careful microscope exams performed before and after the test to verify that the second detector does not have any spots or erosion on the cathode surface capable to provoke sparks. However, as one can see from Fig. 4, the MAC drops with rate identically for both detectors. This demonstrates that the rate effect cannot explained only by possible addition multiplication near the cathode region.

Note that these results presented above were obtained at small $n_0$ (~220 electrons); at larger $n_0$ (larger than $10^4$ electrons) breakdown is certainly dominated by another, space charge, mechanism (see Fig. 5)

Therefore in a real experiment the maximum achievable gain (MAG) will mostly be restricted not by the high-rate events, with small $n_0$, but by heavily ionizing particles.

### *C. Area Effect*

Another important effect is that the MAG drops not only with rate but also with the irradiated area. This effect was first discovered in PPAC [3] and confirmed for GEMs [16], but qualitatively similar results were obtained later for all tested detectors (see Fig. 6). This effect should be taken into account when planning a large-scale experiment.

### *D. Multistep Configurations*

As illustrated in the previous section, the MAC of micropattern detectors is rather limited, especially at small $n_0$. A possibility to exceed this apparent limit is to use one or more steps of multiplication. To study the reasons for this effect we tested most of the micropattern detectors mentioned in section III-A in two configurations: multistep with transfer gaps in between and without them (amplification structures immediately attached to each other).

#### *1) Multistep Configurations with Transfer Gaps*

Fig. 7a shows how the MAC depends on $n_0$ for a number of cascaded detectors. Measurements were done at voltage setting allowing the maximum gain to achieve [17]. One can see that at large $n_0$, in first approximation, the MAC increases linearly with the number of steps and also with the thickness of the transfer gap (see Fig.7b). At small $n_0$ the MAC may significantly increase with the number of steps, up to several orders of magnitude.

At large $n_0$ the MAC depends only slightly on the gas mixture, whereas at small $n_0$ this dependence could be quite strong.

From the results presented in Fig. 2, 3 and 7 one can derive the interesting conclusion that in the first approximation a few steps of "thin" detectors are more or less equivalent to one "thick" detector.

During these studies we also found that the UV photons emitted by the avalanches can play an important role in the operation of multistep detectors. Several experimental works were denote to this subject [17]. We will only briefly mention that at high gas gains avalanche photo-emission may case the following effects: 1) longitudinal spread of the charge cloud entering the transfer region from the detector; 2) discharge propagation from one detector to another one.

The first effect (together with the enhanced diffusion effect [19]) causes a reduction of the ion density in the charge cloud moving into the transfer region, allowing a higher value of the MAC.

The second effect was that, under some conditions, positive ions might "excite" the cathode and cause delayed breakdown [17]. An understanding of the role of the photons and ions provides concrete practical ways for counter design and gas optimisation. For instance, the total gain increased and discharge propagation was suppressed when larger transfer regions were used [17].

*2) Multistep Configurations without Transfer Gaps*

Multistep configurations without transfer regions gave a smaller overall gain and easier discharge propagation. However this type of detector design may be attractive for some applications such as tracking. For example, developed by us MICROMEGAS with preamplification structure [8] are now successfully tested in several experiments and is being considered for tracker applications [20].

During these studies we also found that when the parallel-plate type pre-amplification structure was directly attached to any of the other detectors the position resolution improved remarkably. For instance, a position resolution of 50 μm could be easily obtained directly from the measured analogue signals of the anode strips (without applying any treatment method like the centre of gravity [8], [9]). It is interesting to note however that a comparable position resolution could also be achieved with a large gap RPC (see Fig. 8 and explanation in [8]).

## IV. High Pressure

Detectors operating at high pressure are very attractive for some applications like medical or astrophysical measurements, being also interesting to study micropattern detectors at elevated pressures.

Fig. 9 shows the typical dependence of the MAG on gas pressure. Clearly for all detectors tested the MAG drops strongly with pressure, while the position resolution improves with pressure (Fig. 8).

Earlier results of this type can be found in [18].

## V. Discussion

### A. Very Low Rate

The Raether limit is determined by the charge density in the avalanche and by its ratio with respect to the surface charge density on the detector electrodes. It is not astonishing therefore that at large $n_0$ this limit depends on the detector geometry, gap width, the number of steps, primary electron track length and the gas pressure.

At small $n_0$ the situation is different. In this case, to get the same total charge in avalanche as at large $n_0$, one has to apply the maximum possible voltage to the detector electrodes. Thus the breakdown at small $n_0$ appears due to the combination of several effects: statistical fluctuations in the avalanche final charge $An_0$, imperfections of the detector construction (breakdown along the dielectric surfaces [12] or due to sharp edges, tips, etc), cosmic and ambient radioactivity. Associated to these effects there are also electron jets from the cathode surface [4]. As a result the MAC is smaller than what may be expect from the Raether limit alone.

### B. Rate Effect

As we have already mentioned there are two hypotheses explaining the rate effect at small $n_0$.

The measurements presented in Fig. 4 seem to exclude the hypothesis based on regions of additional gas multiplication near cathodes and therefore indirectly supports alternative explanations, for instance the explanation given in [4]. According to this hypothesis dielectric insertions on the cathode surface may emit jets of electrons. Absorbed layers of some gases or polymer layer due to the aging effects could be also the source of such jets. Accordingly, to optimised detector rate characteristics one should use gases with no polymerisation and no adsorbed layers [4].

### C. Multistep Configurations

At large $n_0$, breakdown due to the space charge mechanism dominates and a possibility to improve the total gain is to use a large gap detector or a multistep configuration. These small gain improvements are due to the enhanced charged cloud diffusion [19] and to the avalanche spread due to photons.

At small $n_0$, however, the increase in MAC can be considerable because each detector step operates at a voltage lower than the possible maximum, reducing many of the deleterious effects mentioned above.

These effects are also particularly sensitive to gas composition, which has a noticeable effect only at small $n_0$.

### D. Comparison with the Large Gap Detectors

Is follows from our data that a multistep micropattern detector configuration is equivalent, in first approximation, to a single thick gap detector. It is therefore interesting to compare micropattern detectors with the traditional large gap detectors: wire and parallel-plate types.

Wire detectors (single or multi) usually do not suffer from destructive sparks, transiting at high gains from an avalanche mode of operation to a Geiger mode or a quenched streamer mode. However, these detectors have reduced time and position resolutions, while RPC's, being also protected against damaging discharges, enjoy an excellent time (50 ps $\sigma$ [21]) and position resolution (30 to 50 μm FWHM [9], [22]), being high counting rates also possible ($10^5$ Hz/mm$^2$ [23]).

### *E. How Could the Micropattern Detectors be Improved?*

Summarizing our results, it follows from Fig. 2 to 7 that at $n_0 > N_{crit}$ no important improvements in the maximum achievable charge are possible, being the maximum achievable charge set by the space charge effect. However some modest improvements are possible by increasing the gap width or by using a few steps of multiplication. By using semiconductive electrodes one can restrict the destructive power of any occasional sparks.

At small $n_0$ the use of the multistep configuration or wide gap detectors is crucial in order to reach high gains. Some additional improvements could be done through the gas (large $d(lnA)/dV$ ) and detector geometry optimisation (geometries insuring fast drop of the electric field with the distance from the anode [12]).

In tracking applications, especially at high rates, it is important to minimize the size of the induced charge region on detector's electrodes in order to improve the position resolution. Unfortunately in the case of the wide gap detectors or multistep configuration the size of the induced signal region increases, but in some applications one can improve this parameter by using a pre-amplification gap.

In general a compromise exists between the composition of the gas mixture, the maximum achievable gain (especially at small $n_0$ ) and the size of the induced charge region. This allows, in some cases, to achieve position resolution better than 30μm in a simple counting mode without use of any treatment methods like centre of gravity [9]. A detailed description of several optimised configurations of micropattern detectors that we successfully used for high rate applications can be found in [8], [9]. As we have already mentioned above, one of them, MICROMEGAS with pre-amplification region, now considered as a tracker detector for several large scale experiments at CERN [20].

## VI. Conclusions

We have described several effects which affect, sometimes strongly, the maximum achievable gain of gaseous detectors: amount of primary charge, total and local counting rate, presence of heavily ionising radiation, number of amplification steps and gas pressure.

The complex and often counterintuitive interplay of all these factors excludes the possibility of an universal optimum solution, applicable to all situations, being however a few guidelines sketched in the discussion.

In some applications it is our opinion that RPCs are excellent. Being protected from damaging discharges by the resistive electrodes, they feature high gain, high rate capability, good position resolution and excellent timing.

## VII. Acknowledgment

We are grateful to F. Sauli and L. Ropelewski for discussions and comments on the manuscript.

## VIII. References

[1] Proceedings of the *International Workshop on Micropattern Detectors,* F. Sauli and M. Lemonnier, Eds. Orsay, France, 1999.

[2] Proceedings of the *PSD99-5th International Conference on Position Sensitive Detectors*, London, 1999 (in press).

[3] Yu. Ivaniouchenkov, P. Fonte, V. Peskov, B.D. Ramsey, "Breakdown limit studies in high- rate gaseous detectors", *Nucl. Instr. and Meth. in Phys. Res. A*, vol. 422, pp. 300-304, 1999.

[4] P. Fonte, V. Peskov, B. Ramsey, "The fundamental limitations of high-rate gaseous detectors", *IEEE Trans. on Nucl. Sci.*, vol. 46, pp. 312-315, 1999.

[5] A. Bressan, M. Hoch, P. Pagano, L. Ropelewski, M. Gruwe, A. Sharma et al., "High rate behavior and discharge limits in micropattern detectors", *Nucl. Instr. and Meth. in Phys. Res. A*, vol. 424, pp. 321-329, 1999.

[6] B. Schmidt, University of Heidelberg, private communication.

[7] R. Bellazzini, INFN, Pisa, private communication.

[8] V. Peskov and P. Fonte, "Gain, rate and position resolution limits of micropattern detectors", in [1], pp. 55

[9] P. Fonte and V. Peskov "Micro-gap parallel-plate chambers with porous secondary electron emitters" *Nucl. Instr. and Meth. in Phys. Res. A*, vol. 454, pp. 260-266, 2000

[10] V. Peskov, E. Silin, T. Sokolova, I. Rodionov, S. Gunji, H. Sakurai "Glass capillary plate - a new high granularity gaseous detector of photons and charged particles *IEEE Trans. on Nucl. Sci* v.47 (2000) 1825.

[11] Gorodezki. Private communication

[12] P. Fonte, V. Peskov, B.D. Ramsey, "Streamers in MSGC's and other gaseous detectors", *ICFA Instrum. Bull.*, vol. 15, Fall 1997 issue, http://www.slac.stanford.edu/pubs/icfa

[13] Yu. Raizer, "Gas discharge physics", Berlin-Heidelberg-New York: Springer-Verlag, 1997.

[14] F. Sauli and L. Ropelewski, CERN, private communication, Montreal 1998.

[15] P. Fonte, V. Peskov, B.D. Ramsey, "Which gaseous detector is the best at high rates", *ICFA Instrum. Bull.*, vol. 16 Summer 1998 issue, http://www.slac.stanford.edu/pubs/icfa

[16] J. Ostling, A. Brahme, M. Danielsson, C. Iacobaeus, V. Peskov "Amplification and conditioning properties of GEM and CAT detector for beam monitoring", in [1], pp. 143.

[17] M. Walmarkt, "Operating range of a GEM for portal imaging", Diploma Thesis, KTH, Stockholm, 2000.
M. Walmarkt et al , "Study of operating limits of a GEM-based portal imaging device", submitted to *Nucl. Instr. and Meth. in Phys. Res. A,*
C.Iacobaeus et al., “A novel portal imaging device for advanced radiation therapy” submitted to the *IEEE Trans. on Nucl. Sci*
J. Ostling et al., “Novel detector for portal imaging in radiation therapy” Proceedings of SPIE-The international Society for Optical Engineering, v. 3977, 2000 p.84

[18] F.A.F. Fraga, M.M.F.R. Fraga, R.F. Marques, L.M.S. Margato, J.R. Gonçalo, A.J.P.L Policarpo, C.W.E. Eijk, R.W. Holander and J. Van der Marel, "Perfomance of microstrip and microgap gas detectors at high pressure," *Nucl. Instr. and Meth. in Phys. Res. A*, vol. 392, pp. 135-138, 1997 [4th International Conference on Position Sensitive Detectors, Manchester, 9-16 September, 1996].

[19] P. Fonte, V.Peskov and B.D. Ramsey, "A study of breakdown limits in microstrip gas counters with preamplification structures", *Nucl. Instr. and Meth. in Phys. Res. A*, vol. 416, pp. 23-31, 1998.

[20] Y.Giomataris, DAPNIA, Saclay, private communication.

[21] P. Fonte, R. Ferreira Marques, J. Pinhão, N. Carolino and A. Policarpo, "High-resolution RPCs for large TOF systems", *Nucl. Instr. and Meth. in Phys. Res. A*, vol. 449, pp. 295-301, 2000.

[22] E. Cerron Zeballos, I. Crotty, P. Fonte, D. Hatzifotiadou, J. Lamas Valverde, V. Peskov et al., "New developments of RPC: secondary electron emission and microstrip readout", *Scientifica Acta*, vol. XI-1, pp. 45-49, 1996.

[23] P. Fonte, N. Carolino, L. Costa, R. Ferreira-Marques, S. Mendiratta, V. Peskov et al., "A spark-protected high rate detector", *Nucl. Instr. and Meth. in Phys. Res. A*, vol. 431, pp. 154-159, 1999.

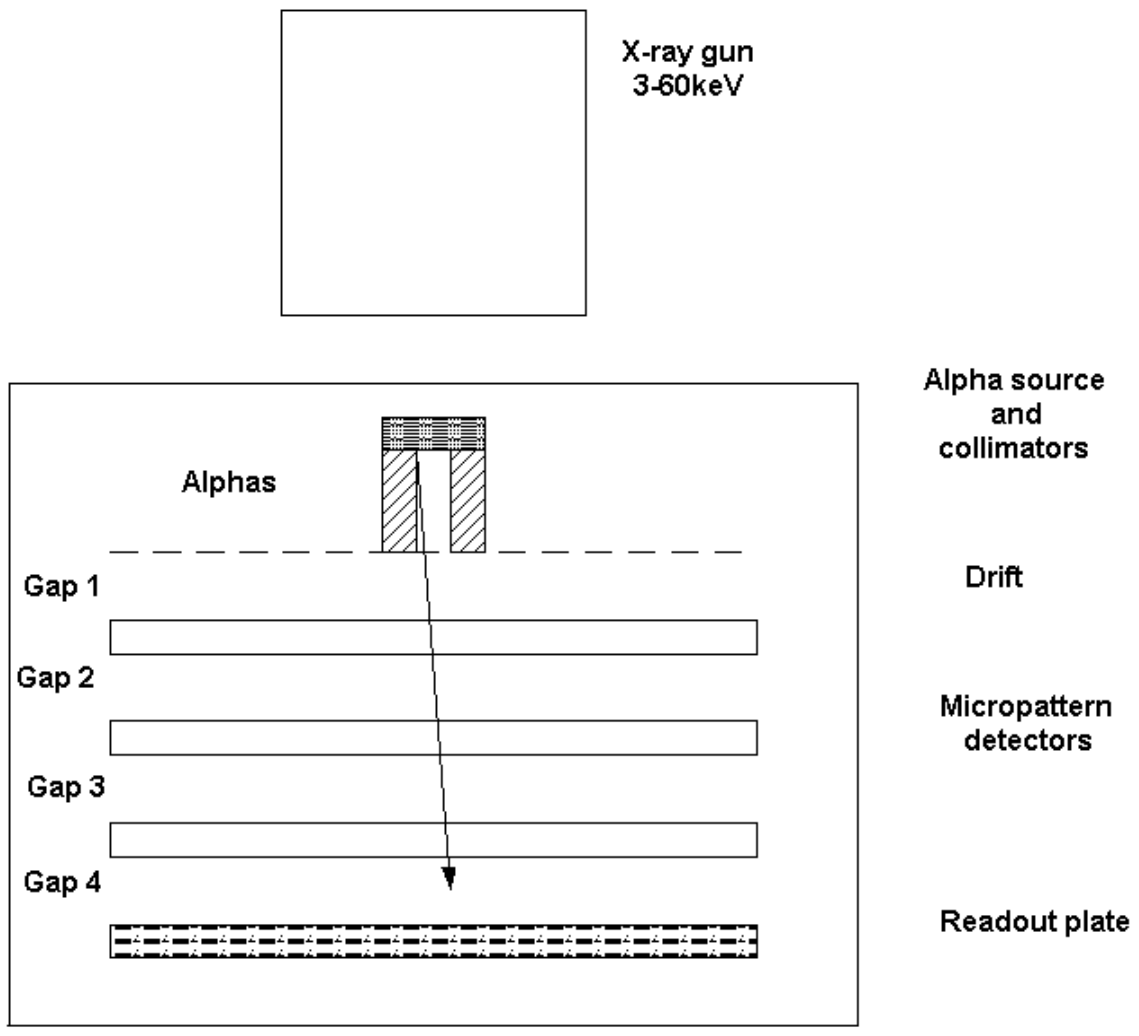

Fig. 1. Schematic drawing of the experimental setup.

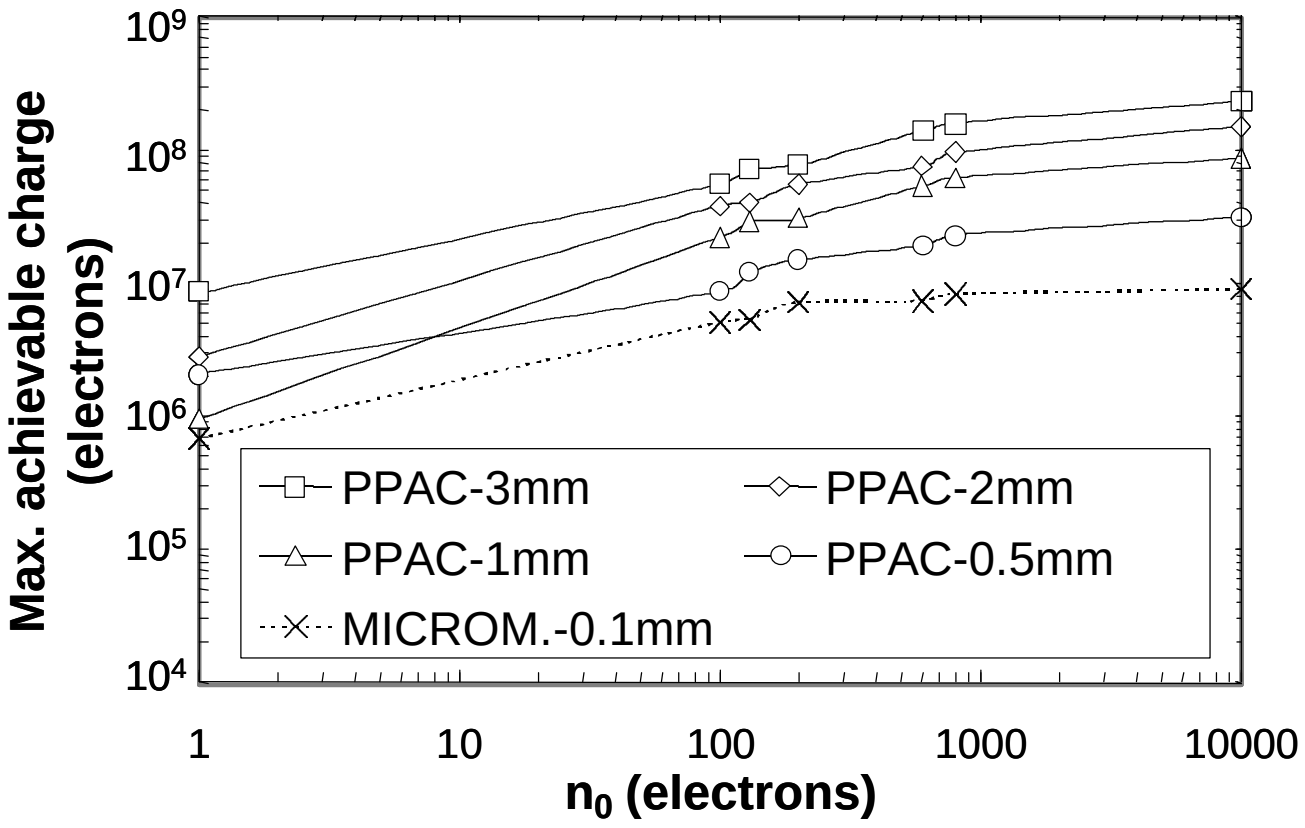

Fig. 2. Dependence of the maximum achievable charge on $n_0$ for PPAC and MICROMEGAS. The measurements with PPAC were done in Ar+10% ethane and with MICROMEGAS in Ar+4%DME at 1 atm.

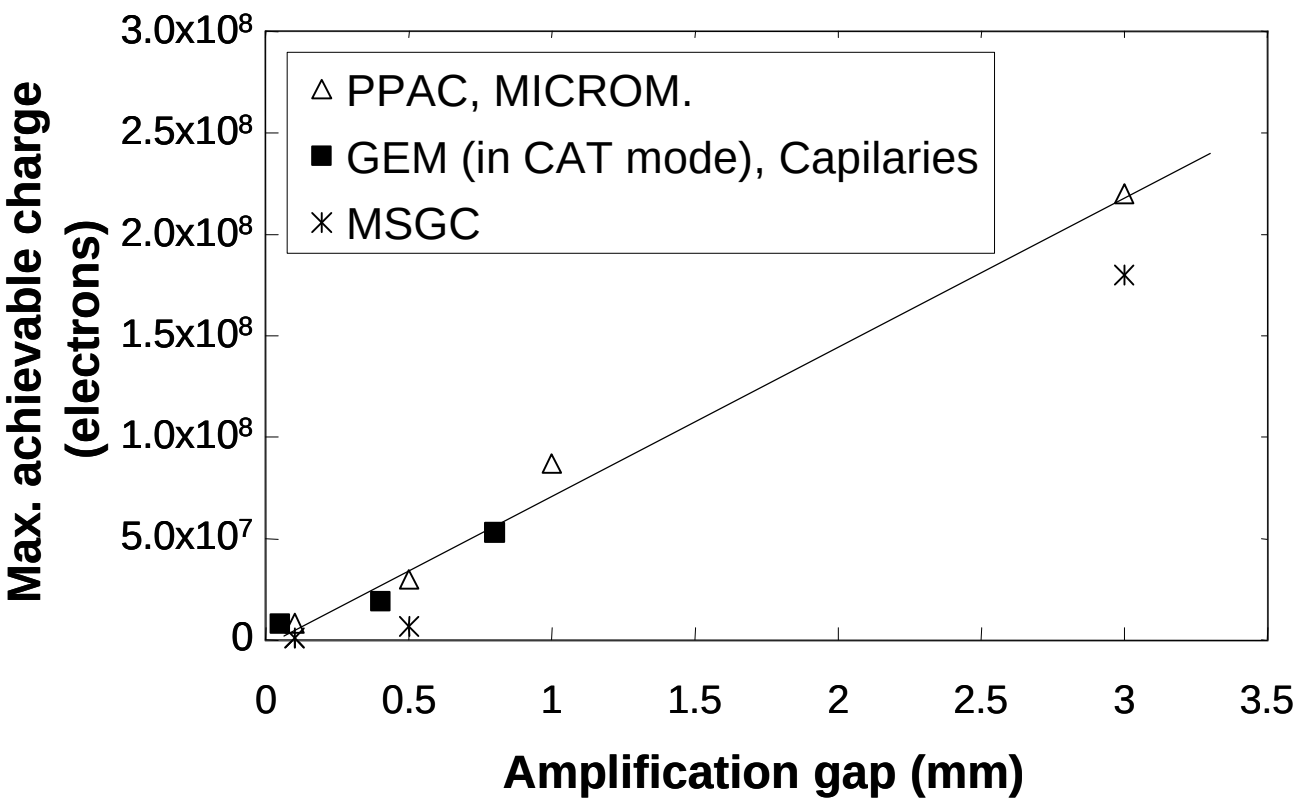

Fig. 3. Dependence of the maximum achievable charge on the amplification gap at $n_0$ □ $10^3$ electrons. The measurements with PPAC were done in Ar+10% ethane, with MSGC in Xe+2% isobutilene, with MICROMEGAS in Ar+4%DME and with GEM and capillaries in Ar+20%$CO_2$. The solid line is a linear fit to the PPAC data.

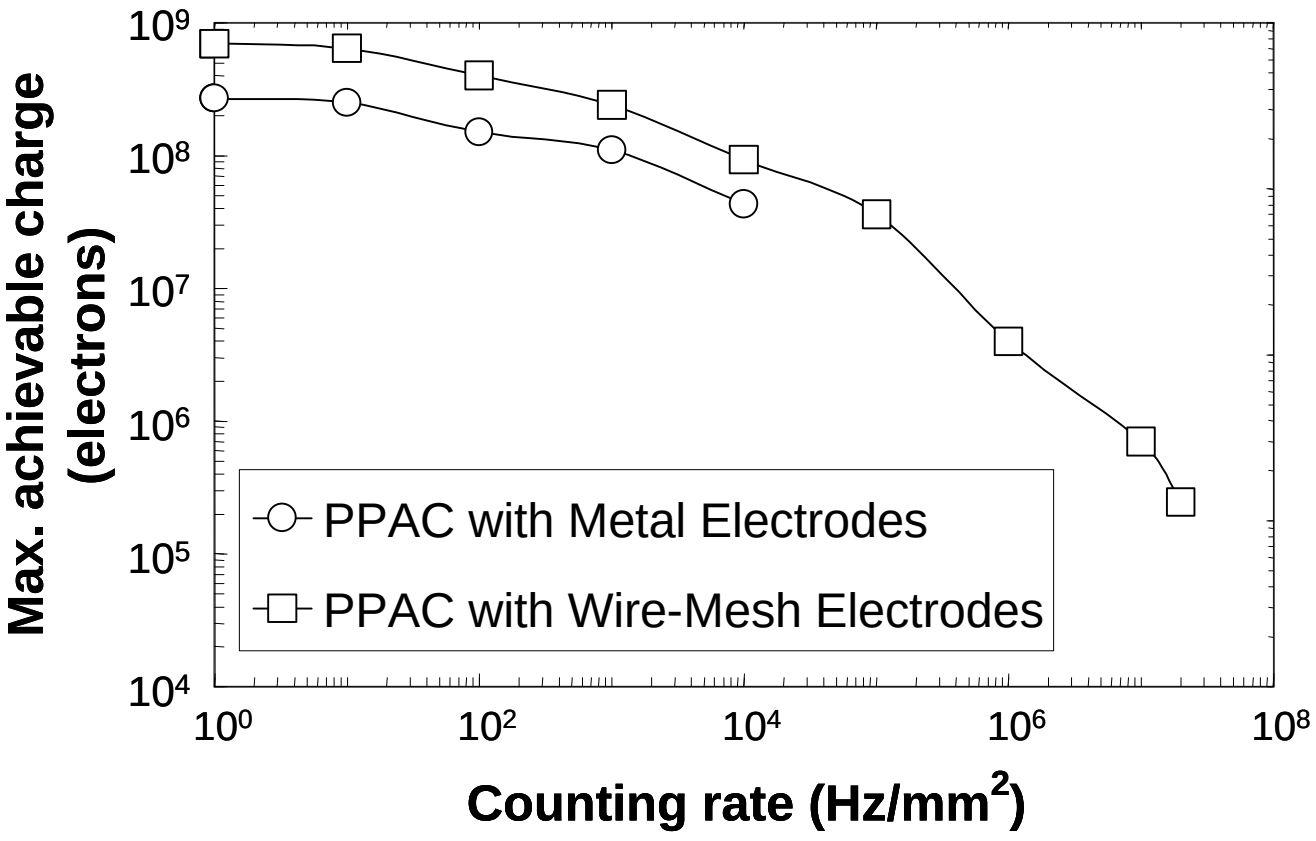

Fig. 4. Maximum achievable charge as function of counting rate for two configurations of PPAC. In both cases the gap was 2 mm and the gas mixture Ar+20% ethane.

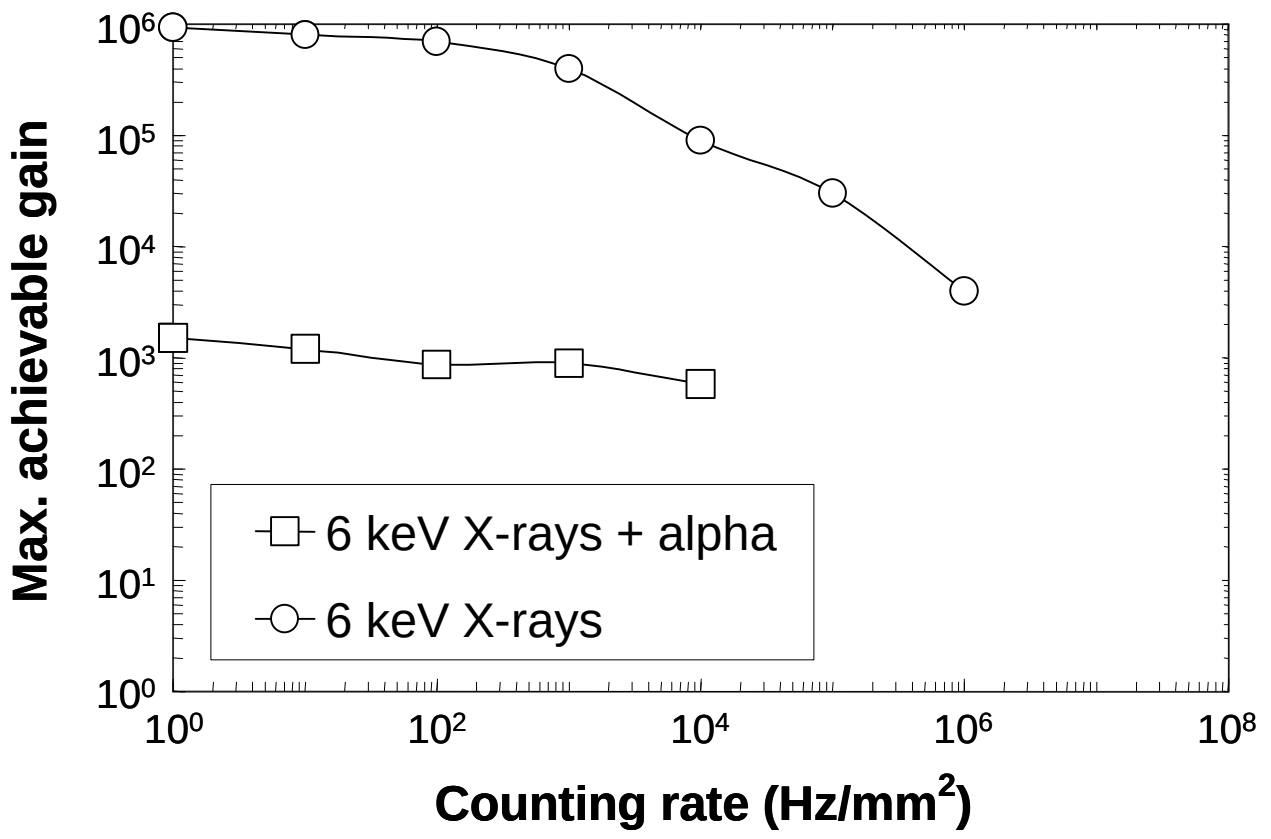

Fig. 5. Maximum achievable gain vs. rate for 6 keV photons alone and in presence of a collimated alpha source (□ $10^3$ Hz). The measurements were made in a PPAC with 3 mm gap and 2 cm drift in a gas mixture of Ar+12% ethane.

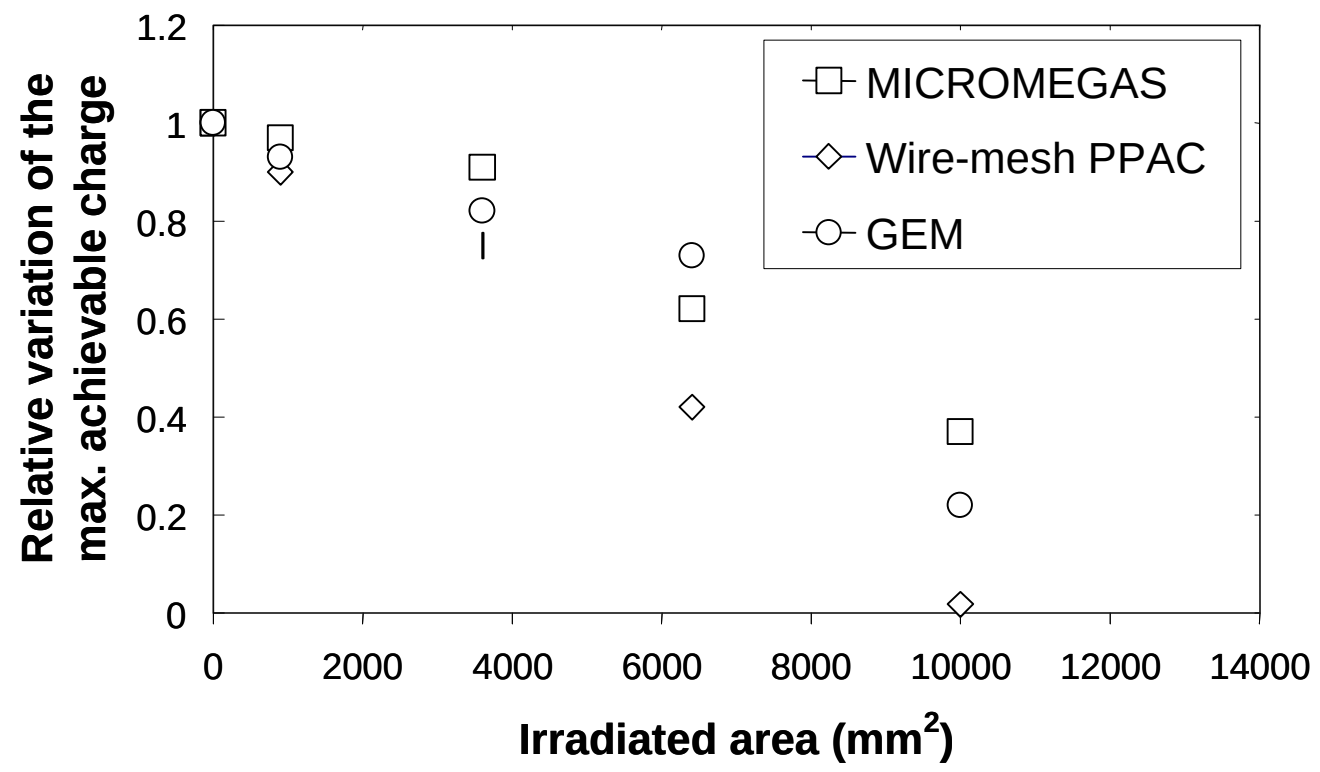

Fig. 6. Relative variation of the maximum achievable charge vs. the irradiated detector area ("area effect"), measured at a flux density of $10^5$ Hz/mm$^2$ of 20 keV X-rays in a gas mixture of Ar+20% $CO_2$.

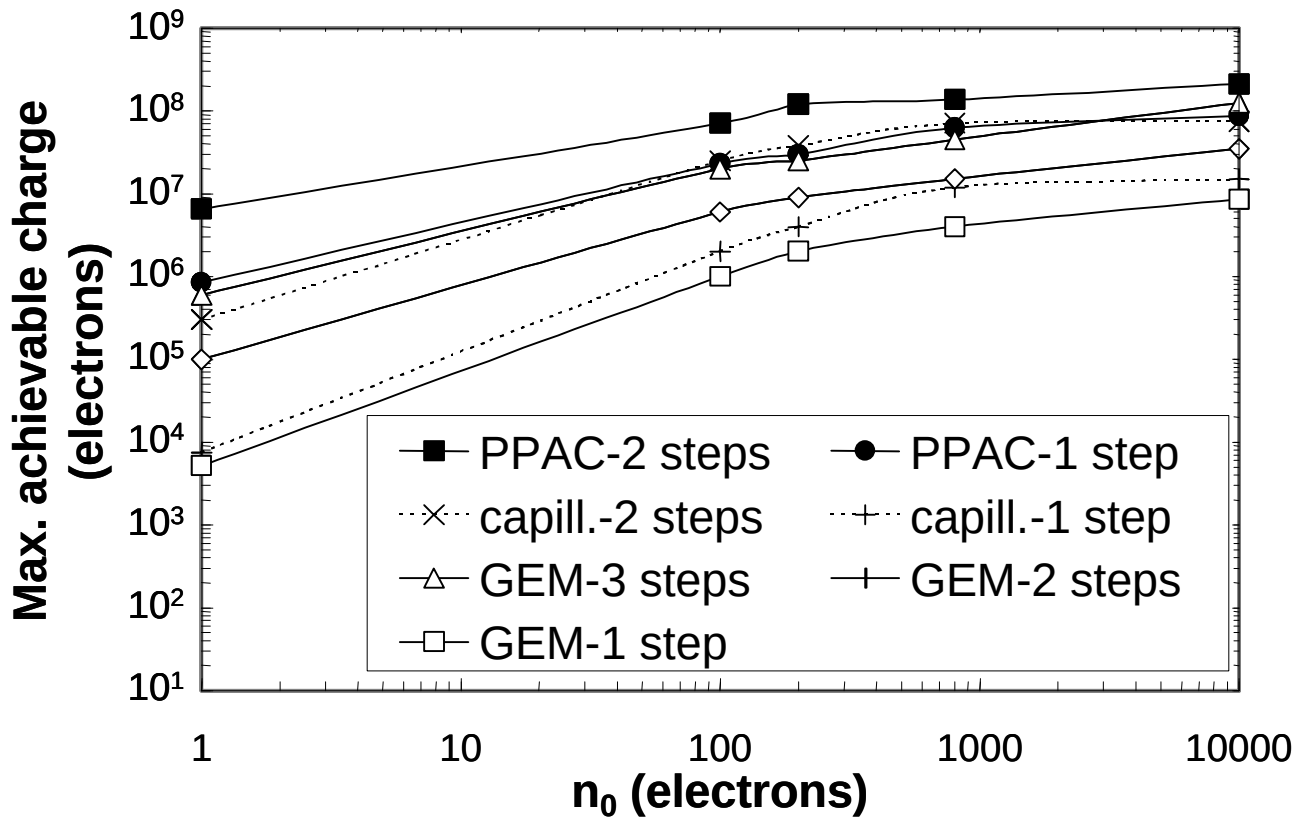

a)

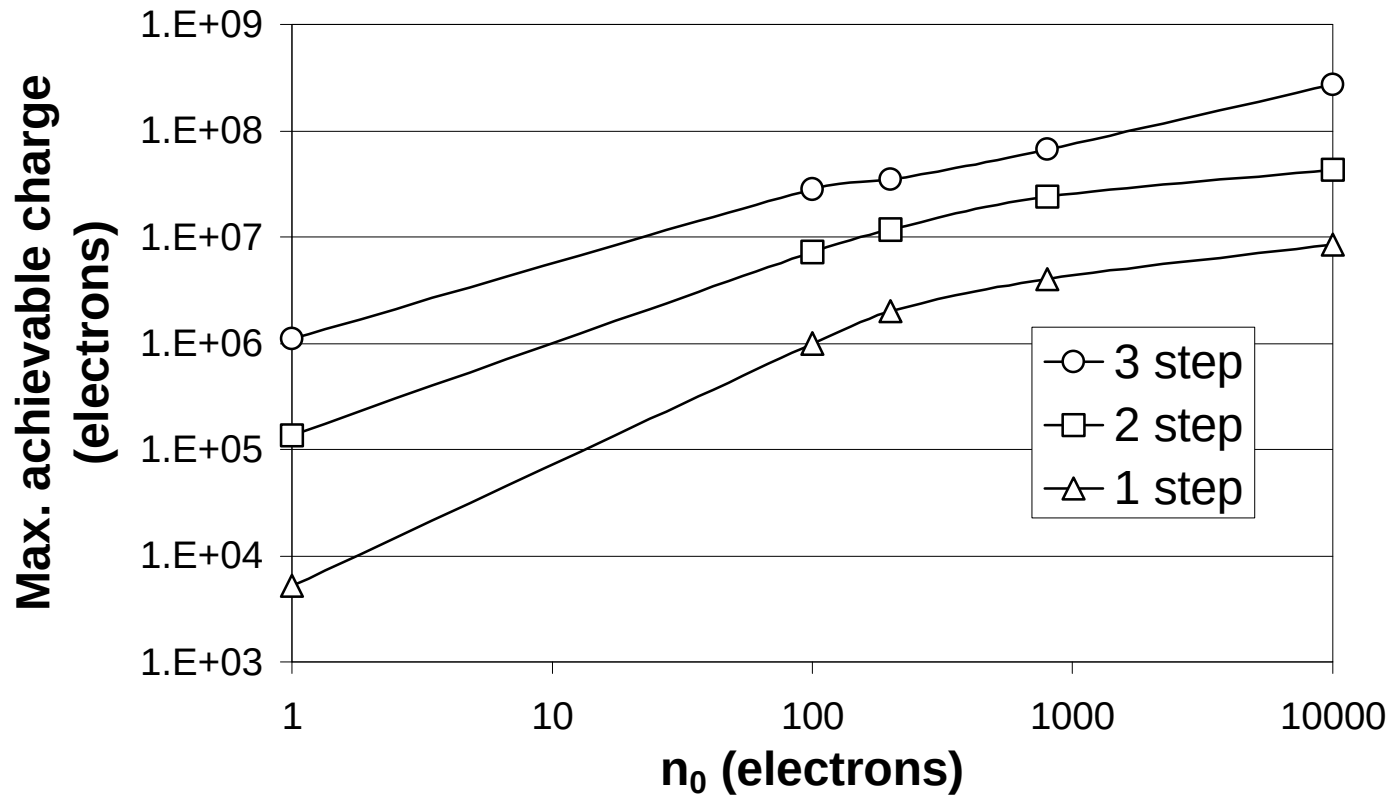

b) .

Fig. 7a. Maximum achievable charge vs. $n_0$ for several detectors, single and multistep: single GEM (in CAT mode), double GEM, triple GEM, capillary plate 0.4 mm thick, double capillary plate, PPAC 1 mm gap, double PPAC. The measurements with PPAC were done in Ar+10% ethane and with GEM and capillaries in Ar+20%$CO_2$. In the case of GEM and capillaries the width of transfer gaps was 1.5 mm. In the case of PPAC it was 2 mm

Fig. 7b. Maximum achievable charge vs. $n_0$ for GEM's with 8 mm transfer gaps

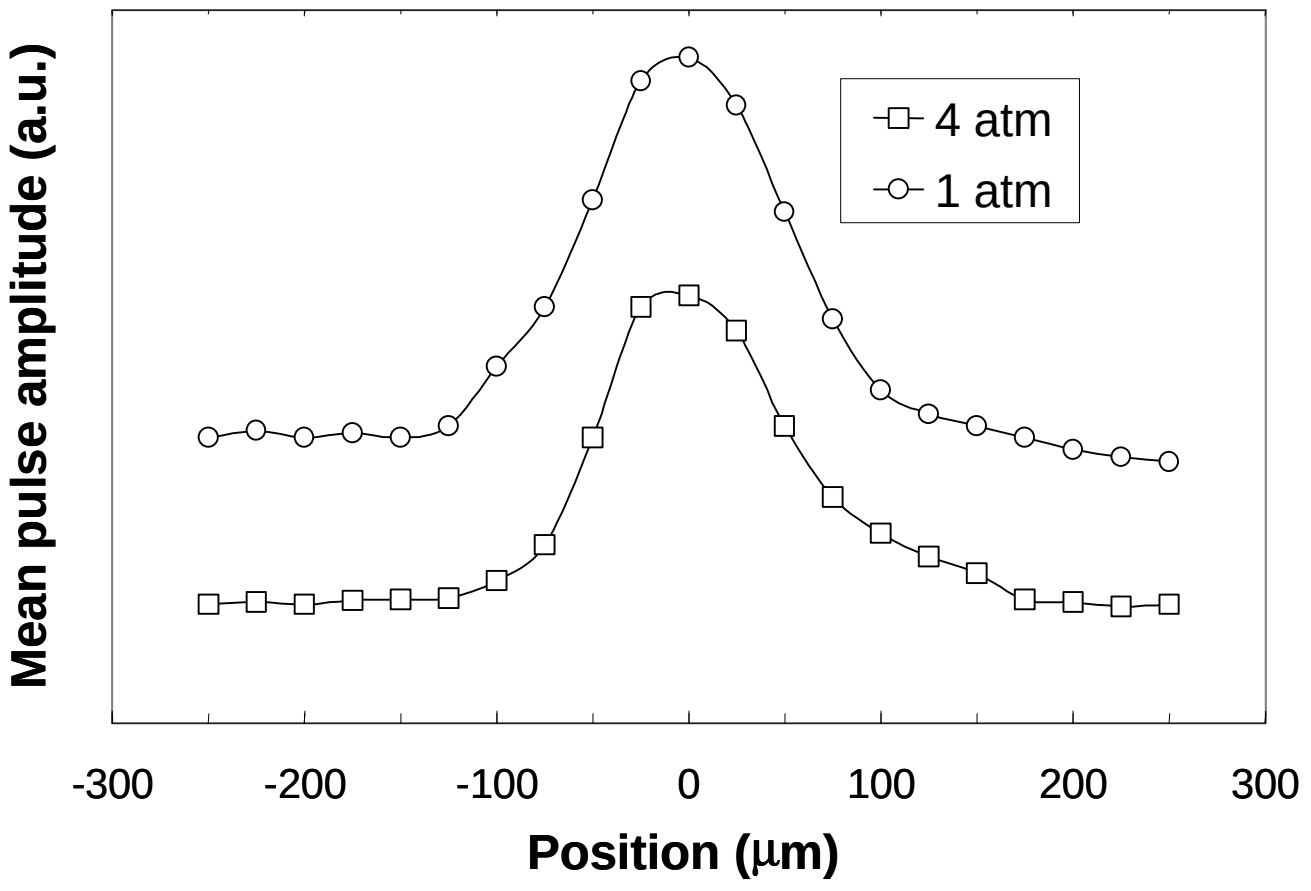

Fig. 8. The mean amplitude of the signals from a 50 μm anode strip vs. position of the 50 μm collimator at 2 different pressures in Kr+20% $CO_2$. The resolution improvement is significant, but small in absolute terms because it is convoluted with the strip and collimator widths.

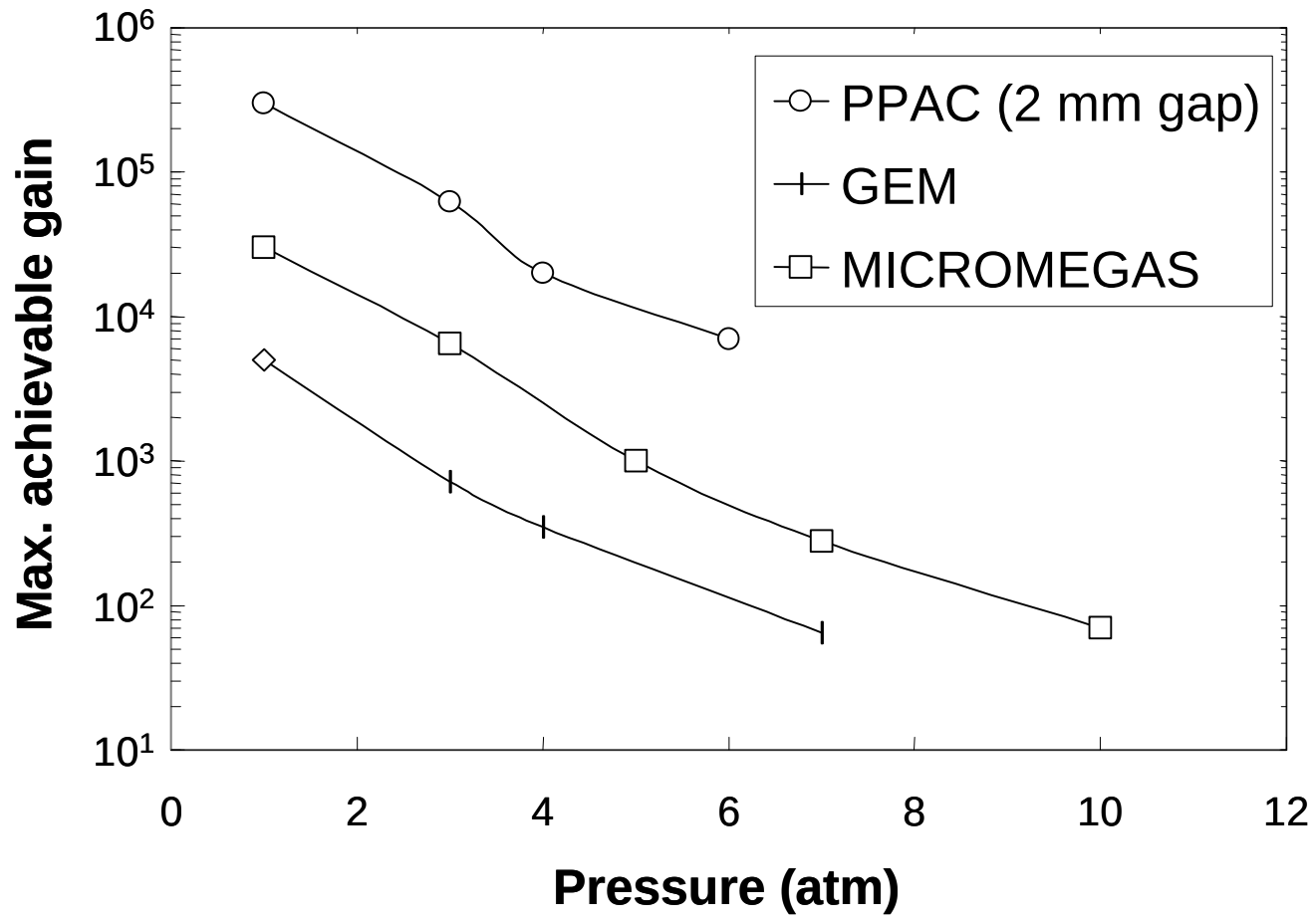

Fig. 9. Typical dependence of the maximum achievable gain as a function of pressure for several detector types. All measurements were done in Xe+40%Kr+$CO_2$.